# Reading ability detection using eye-tracking data with LSTM-based few-shot learning

Nanxi Li, Hongjiang Wang, Zehui Zhan

School of Information Technology in Education, South China Normal University, Guangzhou, 510631, China

pumpkinLNX@gmail.com, wanghongjiang@m.scnu.edu.cn, zhanzehui@m.scnu.edu.cn

**Abstract:**
Reading ability detection is important in modern educational field. In this paper, a method of predicting scores of reading ability is proposed, using the eye-tracking data of a few subjects (e.g., 68 subjects). The proposed method built a regression model for the score prediction by combining Long Short Time Memory (LSTM) and light-weighted neural networks. Experiments show that with few-shot learning strategy, the proposed method achieved higher accuracy than previous methods of score prediction in reading ability detection. The code can later be downloaded at https://github.com/pumpkinLNX/LSTM-eye-tracking-pytorch.git

## 1. Introduction

Reading ability detection is important in modern educational field, which may reveal the subjects' ability of comprehension or cognition during their reading process. Previous works demonstrated that eye-tracking data supplied meaningful information for reading ability detection, and have gained promising results by employing machine learning methods [1-18].

The eye-tracking based methods of reading ability detection fell into two main categories: the one estimated reading ability with finite number of classes [1-14], providing qualitative evaluation of subjects' reading ability. The other predicted reading ability scores with regression models [15-18], rendering quantitative evaluation of subjects' reading ability. Although the former exhibited satisfactory accuracy in detecting certain classes of abnormalities in reading, it lacked the capability of predicting exact scores of reading ability, which was emphasized in highly interactive educational environments (such as online learning) to make personal and intelligent reactions to subjects.

However, precise score prediction of reading ability using eye-tracking data is not easy [15-18], especially when the sample data of subjects are few. In this paper, with few-shot learning strategy, a regression model for score prediction is proposed by combining Long Short Time

Memory (LSTM) [19] and light-weighted neural networks. The proposed model exhibits higher accuracy than previous score prediction models tested on the same dataset.

## 2. Related works

Previous eye-tracking based reading ability detection methods have showed great interest in detecting finite classes of abnormalities [1- 14]. For dyslexia detection, Nerušil et al [1] adopted a Convolutional Neural Network (CNN) to classify entire eye-tracking records either in time or frequency as a whole. D'Mello [2] utilized random forest models, VAJS et al [6] employed autoencoder neural network, Iaconis et al [8] used ordinal pattern transition networks, Hmimdi et al [4] presented a convolutional network tested on a large dataset, Nagarajan et al [9] introduced an explainable reinforcement learning model, and Qi et al [12] proposed a temporal convolutional network (called CASES), respectively, to identify abnormalities in reading. Vajs et al [5] tested new features of eye tracking data on four types of classifiers and compared the results for dyslexia detection. Yoo [10] proposed a quantitative tool to diagnose reading disorders with seven machine learning models for binary classification. Prasse et al [11] developed a neural network SP-EyeGAN to generate synthetic raw eye tracking data, and found that pre-training on the synthetic data was beneficial for native reader classification. Xia et al [14] employed an interpretable model (namely SHAP) to explain the effects of eye tracking features in English proficiency level classification. For dyslexia detection among Chinese children, Haller [3] proposed sequential models to process eye movements and performed tests on Mandarin Chinese. Shi et al [13] used support vector machine to classifier subjects' Chinese reading proficiency levels. Apart from traditional eye trackers, webcam based eye tracker was employed by Hutt et al [7] to test low-cost eye tracking. All the above methods revealed different aspects of eye-tracking based reading ability classification and have achieved satisfactory classification accuracy.

One the other hand, research on score prediction of reading ability using eye-tracking data was relatively few. In previous works, linear regression models [15-17] were usually adopted to predict scores of reading comprehension on English datasets. Besides, Zhan et al [18] proposed a Multi-Trial Joint Learning Model (MTLM) with linear regression for score prediction on a Chinese dataset. However, due to the few number of eye-tracking samples used in these methods, more complex machine learning methods were hard to be applied for a more accurate score predictor.

Few-shot learning methods [20-25] that occurred in recent years may address the problem mentioned above. Iwata et al [26] proposed a few-shot learning method for spatial regression, using neural networks with Gaussian processes framework. For vision regression tasks such as pose regression, Gao et al [27] presented an addition of Functional Contrastive Learning (FCL) over the task representations in Conditional Neural Processes (CNPs). For tasks of both few-shot classification and few-shot regression, Baik et al [28] designed a framework Meta-Learning with Task-Adaptive Loss Function (MeTAL) for meta-learning with adaptive loss function, Chen et al [29] proposed a nonparametric method in deep embedded space to address incremental few-shot learning, and Tian et al [30] presented a consistent meta-regularization (Con-MetaReg) in meta-learning models to help reduce the data-distribution discrepancy between training data and testing data. Moreover, Satrya et al [31] proposed an ensemble scheme combing Model-Agnostic Meta-Learning (MAML) and transfer learning, which was tested on three regression problems and achieved good performance. Baik et al [32] designed a fast adaptation process for MAML in diverse few-shot learning tasks including classification, regression, tracking and interpolation. For runoff prediction in data scarce regions, Yang et al [33] proposed a LSTM-prototypical network fusion model based on few-shot learning method. Huisman et al [34] studied on LSTM for few-shot learning tasks and found that LSTM outperformed the popular meta-learning technique MAML on simple few-shot regression tasks.

In this paper, a LSTM-based few-shot learning method is proposed for score prediction of reading ability using eye-tracking data. The contribution consists of the following three aspects: (1) Unlike previous methods that used raw eye-tracking data for linear regression, we embedded eye-tracking data into a new vector space using LSTM, expecting to utilize the data dependencies among a group of subjects in the same reading test. (2) Few-shot learning strategy was employed by the proposed method to alleviate the problem of lacking training data. (3) Instead of manually designed formulas for parameter estimation of linear regression in previous methods, light-weighted neural networks were designed for automatic parameter estimation of linear regression in the proposed method.

## 3. Methodology

The framework of the proposed method is shown in Figure 1. Given the eye-tracking data $\vec{X}_i$ of a subject in one reading test, and a group of eye-tracking data $\{\vec{X}_{i-1}, \vec{X}_{i-2}, ......, \vec{X}_{i-(k-1)}\}$ of certain subjects in the same reading test, LSTM is employed to embed the raw vector $\vec{X}_i$ into a new vector $\vec{Z}_i$, which is later used in linear regression for score prediction $y_i^*$ of the reading ability of the subject. This process is indicated by the solid lines and the solid line boxes in Figure 1, which reveals the testing phase of the proposed method.

As for the training phase of the proposed method, both solid and dashed lines / boxes in Figure 1 are employed. Given a subject's eye-tracking data $\vec{X}_i$ in one reading test with its label $y_i$, and a group of eye-tracking data $\{\vec{X}_{i-1}, \vec{X}_{i-2}, ......, \vec{X}_{i-(k-1)}\}$ of certain subjects in the same reading test with their labels $\{y_{i-1}, y_{-2}, ......y_{i-(k-1)}\}$, the data and labels are split into four streams. Then by using LSTM, the data streams $\vec{X}_i$ and $\{\vec{X}_{i-1}, \vec{X}_{i-2}, ......, \vec{X}_{i-(k-1)}\}$ are utilized to embed the vector $\vec{X}_i$ into a new vector $\vec{Z}_i$, as well as to embed the vectors $\{\vec{X}_{i-1}, \vec{X}_{i-2}, ......, \vec{X}_{i-(k-1)}\}$ into new vectors $\{\vec{Z}_{i-1}, \vec{Z}_{i-2}, ......, \vec{Z}_{i-(k-1)}\}$. Subsequently, by adopting light-weighted neural networks, the data stream $\{y_{i-1}, y_{i-2}, ......, y_{i-(k-1)}\}$ is combined with the embedded vectors $\{\vec{Z}_{i-1}, \vec{Z}_{i-2}, ......, \vec{Z}_{i-(k-1)}\}$ to estimate the parameters $\{\vec{w}, \beta\}$ which will be used in linear regression. Next, the embedded vector $\vec{Z}_i$ is fed into linear regression for score prediction $y_i^*$, which is later compared with the ground-truth label $y_i$ for loss computation. Finally, the computed loss is fed back to update the parameters of both LSTM and light-weighted neural networks. The above process in training phase is repeated until an optimal model for score prediction is achieved.

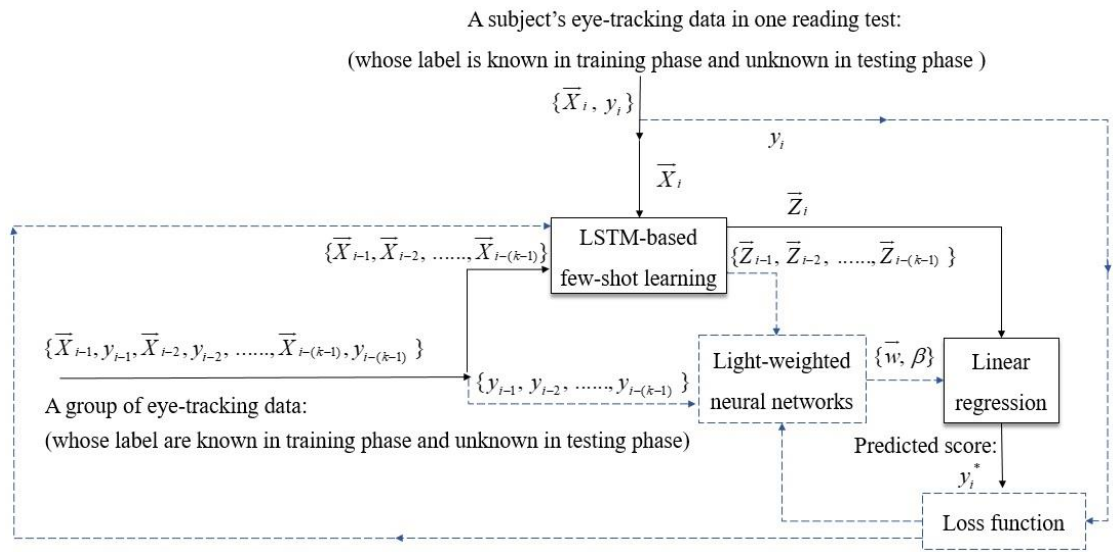

Figure 1 The framework of the proposed method.

3.1 LSTM-based few-shot learning

The LSTM-based few-shot learning module carries out two types of tasks: one is the vector embedding for raw eye-tracking data using LSTM, the other is the few-shot learning strategy combined with the vector embedding.

### 3.1.1 Vector embedding using LSTM

The principle of LSTM is demonstrated in [19, 35], and we utilized the hidden units in LSTM for our vector embedding, as is described below:

$$\{\vec{Z}_i, \vec{Z}_{i-1}, \vec{Z}_{i-2}, ...... \vec{Z}_{i-(k-1)}\} = f_{hidden}(\{\vec{X}_i, \vec{X}_{i-1}, \vec{X}_{i-2}, ...... \vec{X}_{i-(k-1)}\}) \quad (1)$$

where $\vec{X}_i$ is the subject's eye-tracking data in one reading test, $\{\vec{X}_{i-1}, \vec{X}_{i-2}, ......, \vec{X}_{i-(k-1)}\}$ is the raw eye-tracking data of a group of subjects in the same reading test, and the sequence $\{\vec{X}_i, \vec{X}_{i-1}, \vec{X}_{i-2}, ......, \vec{X}_{i-(k-1)}\}$ forms the input sequence of LSTM. Similarly, the sequence $\{\vec{Z}_i, \vec{Z}_{i-1}, \vec{Z}_{i-2}, ......, \vec{Z}_{i-(k-1)}\}$ forms the hidden state sequence output by LSTM. Besides, the function $f_{hidden}$ represents the operation of producing hidden state sequence by LSTM, which can be inferred from the following equations described in LSTM [19, 35]:

$$\overrightarrow{forget}_i = \sigma(W_f * [\vec{Z}_{i-1}, \vec{X}_i] + \vec{B}_f) \quad (2)$$

$$\overrightarrow{gate}_i = \tanh(W_g * [\vec{Z}_{i-1}, \vec{X}_i] + \vec{B}_g) \quad (3)$$

$$\overrightarrow{in}_i = \sigma(W_i * [\vec{Z}_{i-1}, \vec{X}_i] + \vec{B}_i) \quad (4)$$

$$\overrightarrow{out}_i = \sigma(W_o * [\vec{Z}_{i-1}, \vec{X}_i] + \vec{B}_o) \quad (5)$$

$$\overrightarrow{cell}_i = \overrightarrow{cell}_{i-1} \circ \overrightarrow{forget}_i + \overrightarrow{gate}_i \circ \overrightarrow{in}_i \quad (6)$$

$$\vec{Z}_i = \overrightarrow{out}_i \circ \tanh(\overrightarrow{cell}_i) \quad (7)$$

where $W_f$, and $\vec{B}_f$ are respectively the weight and bias for the forget gate in LSTM, $W_g$ and $W_i$ are the weights for the input gate of LSTM, $\vec{B}_g$ and $\vec{B}_i$ are the biases for the input gate of LSTM, $W_o$, and $\vec{B}_o$ are respectively the weight and bias for the output gate of LSTM. $\vec{X}_i$ is the raw vector 'currently' input into LSTM, and $\vec{Z}_{i-1}$ is the hidden state vector produced by LSTM which corresponds to the 'historical' raw vector $\vec{X}_{i-1}$. The functions $\sigma$ and $\tanh$ are sigmoid function and hyperbolic tangent function, respectively. And the operators * and ∘ are matrix multiplication and Hadamard product, respectively.

As can be seen in equation (7), LSTM outputs the 'current' hidden state vector $\vec{Z}_i$ which corresponds to the raw vector $\vec{X}_i$. It forms the hidden state sequence $\{\vec{Z}_i, \vec{Z}_{i-1}, \vec{Z}_{i-2}, ......, \vec{Z}_{i-(k-1)}\}$ along with the 'historical' hidden state vectors $\{\vec{Z}_{i-1}, \vec{Z}_{i-2}, ......, \vec{Z}_{i-(k-1)}\}$ that are sequentially produced by LSTM. We treat this hidden state sequence $\{\vec{Z}_i, \vec{Z}_{i-1}, \vec{Z}_{i-2}, ......, \vec{Z}_{i-(k-1)}\}$ as the embedding vector sequence of the raw vector sequence $\{\vec{X}_i, \vec{X}_{i-1}, \vec{X}_{i-2}, ......, \vec{X}_{i-(k-1)}\}$, and the whole process is represented in equation (1) for simplicity.

The above process of vector embedding using LSTM might be beneficial for representing subjects' reading abilities with lower number of features (as the vector $\vec{Z}_i$ often has much lower dimension than vector $\vec{X}_i$) while capturing the dependencies among different subjects (due to the memorization characteristic of LSTM). As far as we know, the proposed method is a first attempt to apply vector embedding to score prediction of reading ability using LSTM.

3.1.2 Few-shot learning strategy

We utilized the idea of episodic training [20, 26, 36-38] to implement our few-shot learning strategy. Since episodic training differs from classical training in that for the former, each 'episode' in training phase is itself a training-and-testing process, we organized the data for our episodic training as is illustrated in Figure 2, and designed our episodic training strategy as is described in Algorithm 1.

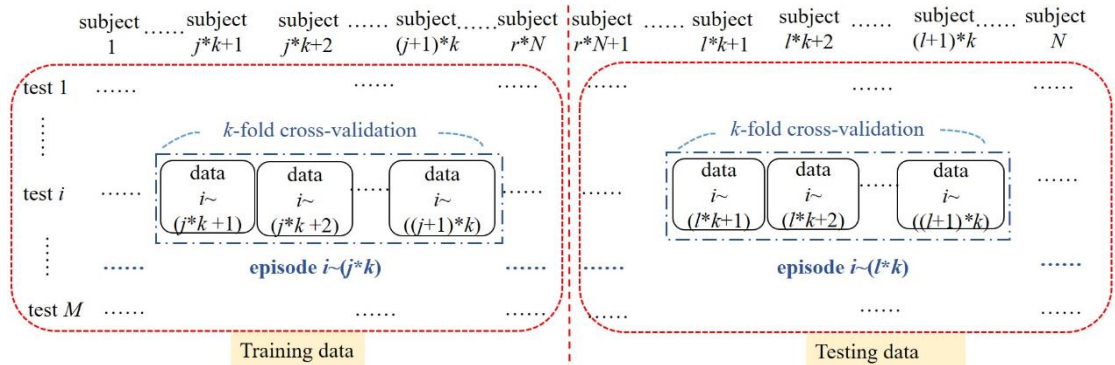

Figure 2 Data organization for the episodic training of the proposed method

As is illustrated in Figure 2, the data for episodic training in the proposed method is organized as following: Given $N$ subjects each taking $M$ reading tests, we firstly extracted the training data by multiplying the number of reading subjects $N$ with a factor $r$ ($0 < r < 1$), and the eye-tracking data of randomly selected $r*N$ subjects were regarded as training data. Consequently, the rest of data were treated as testing data. Then the $M$ tests for each subject was also randomly arranged. And for each test $i$ ($1 \leq i \leq M$), the eye-tracking data of $k$ non-overlapping neighboring subjects were chosen as an episode (in Figure 2, $j$ and $l$ are both integers in range (0, $N/k$)). For each episode in either training data or testing data, $k$-fold cross-validation could be performed if necessary, which will be detailed in in Section 3.4.

**Algorithm 1** Few-shot learning strategy of the proposed method

**Input:** Organized eye-tracking data in episodes (as is illustrated in Figure 2)

**Training phase**:

1. **For** each episode **in** training data**:**

2. Split the episode $\{\vec{X}_i, y_i, \vec{X}_{i-1}, y_{i-1}, \vec{X}_{i-2}, y_{i-2}, ......, \vec{X}_{i-(k-1)}, y_{i-(k-1)}\}$ into four streams, as is illustrated in Figure 1

3. Input the vector sequence $\{\vec{X}_i, \vec{X}_{i-1}, \vec{X}_{i-2}, ......, \vec{X}_{i-(k-1)}\}$ to LSTM

4. Perform vector embedding using LSTM as is described in equation (1)

5. Estimate the parameters for linear regression using light-weighted neural networks as is demonstrated in Section 3.2

6. Compute the predicted score $y_i^*$ using linear regression.

7. Compare the predicted score and the ground-truth score, and calculate the loss according to equation (8)

8. Update the parameters of LSTM and light-weighted neural networks using back propagation with the gradients of the loss

9. Carry out *k*-fold cross-validation for the input episode if necessary, as is described in Section 3.4

10. **End for**

**Output**: The parameters of trained LSTM and light-weighted neural networks

**Testing phase:**

1. **For** each episode **in** testing data**:**

2. Input the vector sequence $\{\vec{X}_i, \vec{X}_{i-1}, \vec{X}_{i-2}, ......, \vec{X}_{i-(k-1)}\}$ to LSTM

3. Perform vector embedding using LSTM as is described in equation (1)

4. Calculate the predicted score $y_i^*$ for the episode using linear regression

5. Carry out *k*-fold cross-validation for the input episode if necessary, as is described in Section 3.4

6. **End for**

**Output**: The predicted scores of reading ability for testing data

3.2 Light-weighted neural networks

Instead of using manually designed formulas for parameter estimation of linear regression, like in previous works, we devised light-weighted neural networks to automatically estimate the parameters of linear regression, which had the architecture shown in Figure 3:

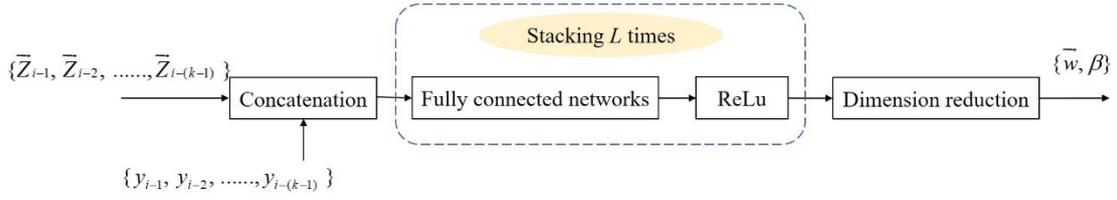

Figure 3 The architecture of the light-weighted neural networks in the proposed method

In the proposed light-weighted neural networks, firstly each label $y_s\ (i-1 \leq s \leq i-(k-1))$ is appended to its corresponding embedded vector $\vec{Z}_s\ (i-1 \leq s \leq i-(k-1))$ to form a new vector, so that the new vector has the dimension more than that of $\vec{Z}_s$ by one. Then this new vector is fed into a stacking structure, where the combination of fully connected networks and ReLu function stacks for $L$ times. For the output of the stacking structure, dimension reduction is carried out to turn the output matrix into a vector which contains the needed parameter $\{\vec{w}, \beta\}$ of linear regression. In the proposed method, the operation of dimension reduction is simply implemented by adding the elements of a matrix along the row dimension.

3.3 Loss function

For training the parameters of the proposed method, we employed the classic Mean Square Error (MSE) loss function below:

$$loss_{MSE} = \frac{1}{n} * \sum_{i=1}^{n}(y_i - y_i^*)^2 \qquad (8)$$

where $y_i^*$ and $y_i$ are the predicted score and the ground-truth score of eye-tracking data $\vec{X}_i$, respectively. And $n$ is the number of episodes in a training batch.

3.4 *k*-fold cross-validation

Since given each episode containing the eye-tracking data of *k* subjects, we predict the score of the reading ability of just one subject, leaving the score prediction for the rest (*k*-1) subjects not addressed. To better utilize either training data or testing data, *k*-fold cross-validation could be employed in our few-shot learning strategy as following: In each episode, the score prediction is carried out in turn for every one of the *k* subjects, i.e., any subject in an episode would have its score prediction using its own eye-tracking data along with those of the rest (*k*-1) subjects in the same episode, through circularly shifting the data sequence in an episode by one subject each time (in total *k* times of circular shift for all subjects in an episode).

**4. Experimental results**

We carried out experiments on the same dataset as in [18], where 74 subjects each taking 42 reading tests were involved in collecting eye-tracking data. After filtering out singular values such as missing features, incomplete test records, etc., we finally used 68 subjects each taking 42 reading tests in our experiments. We set the factor $r$ of extracting training data to be 0.9, the number $k$ of neighboring subjects in an episode to be 3, and the number $L$ of stacking times in light-weighted networks to be 4. The following criterion of regression were employed to evaluate the model performance: Mean Absolute Error (MAE) along with the Standard Deviation (SD), which was the same as that in [18]. (Note that the MAE in Table 1 was denoted by percentage which was similar in [18], e.g., 4.02% meant that the MAE was 4.02 out of 100.)

To find out whether or not $k$-fold cross-validation is helpful to the proposed method, we tested the proposed method under several different configurations, as is shown in Table 1. Two types of feature configurations were employed in our experiments: One was the 19-dimensional features and the other was the 22-dimensional features, both were exploited by [18]. The difference between these two feature configurations is that the latter added 3 indicators to the former, i.e., the scores of Chinese, mathematics, and English in university entrance examination. For simplicity, we noted as N/A for the case where neither training phase nor testing phase had $k$-fold cross-validation, Semi for the case where only training phase took $k$-fold cross-validation, and Full for the case where both training phase and testing phase enjoyed $k$-fold cross-validation.

Table 1 The proposed method under different configurations

| The proposed method with **19** features | | | | | The proposed method with **22** features | | | | |
|---|---|---|---|---|---|---|---|---|---|
| Conf. | Training | | **Testing** | | Conf. | Training | | **Testing** | |
| | MAE | SD | MAE | SD | | MAE | SD | MAE | SD |
| N/A | 5.46% | 0.0417 | 9.01% | 0.0547 | N/A | 4.30% | 0.0324 | 6.66% | 0.0620 |
| Semi | 5.20% | 0.0412 | 6.56% | 0.0395 | Semi | 3.67% | 0.0294 | 4.12% | 0.0333 |
| **Full** | 5.20% | 0.0412 | **6.50%** | 0.0384 | **Full** | 3.67% | 0.0294 | **4.02%** | 0.0344 |

From table 1, we can see that 22 features outperform 19 features in the proposed method, mainly due to the fact that the 3 extra indicators in the former provide meaningful information for the score prediction of reading ability. And we also find that the 'Full' configuration works best for the proposed method with either 19-dimensional features or 22-dimensional features, showing the effectiveness of $k$-fold cross-validation for the proposed method. And we still note that under the 'Semi' configuration, both training and testing accuracy are very similar to those of 'Full' configuration, but much better than those of 'N/A' configuration, indicating that *k*-fold cross-validation works more effectively in training phase than in testing phase.

We then compared our episodic training strategy (for few-shot training) with traditional training strategy. Note that for traditional training each subject in training data is embedded with its (*k*-1) neighbors by LSTM. The results of this comparison is listed in Table 2. For simplicity, we used 'TT' to represent traditional training and 'ET' for episodic training, respectively.

Table 2 Comparison between traditional training and the episodic training strategy

| Conf. | The proposed method with **22** features | | | |
|---|---|---|---|---|
| | Training | | **Testing** | |
| | MAE | SD | MAE | SD |
| TT | **2.06%** | 0.0178 | 6.88% | 0.0465 |
| **ET** | 3.67% | 0.0294 | **4.02%** | 0.0344 |

In Table2, we can find that although presenting larger MAE in training phase, our episodic training strategy outperforms traditional training in testing phase with much lower MAE, showing its effectiveness. This is mainly due to the fact that the episodic training can alleviate the overfitting problems encountered in traditional training when training data are few, so providing better model generality for new data.

For completeness, examples of score prediction under 'Full' configuration with 22 features using episodic training strategy are demonstrated in Figure 4. The scores of reading ability is normalized to range [0, 1] via dividing the original score in range [0, 100] by a constant 100, and are illustrated by 'training data' and 'testing data' in Figure 4. And the training loss and testing loss are calculated based on the normalized scores, as is illustrated in Figure 4.

From Figure 4, we can see that the predicted data can almost cover the ground truth data, both in training phase and testing phase, indicating the precision of the score prediction of the proposed method. On the other hand, the loss values in both training phase and testing phase are quite small, implying the high accuracy of the score prediction of the proposed method.

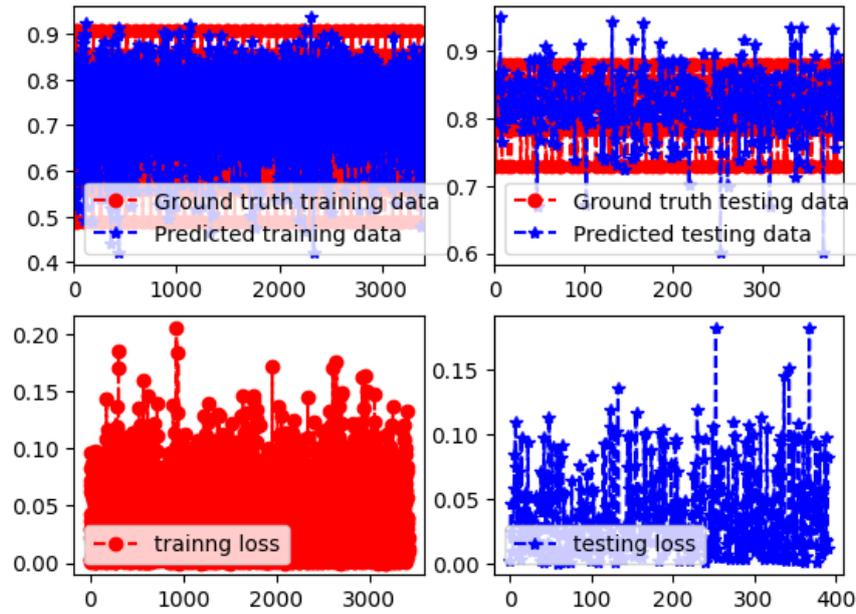

Figure 4 Examples of score prediction under 'Full' configuration with 22 features using episodic training

Furthermore, we analyze the impact of each of the 22 features [18] on the accuracy of the proposed method using Shapley Additive Explanations (SHAP) library [39- 40], which is a powerful tool for explaining the feature impacts of various models in artificial intelligence.

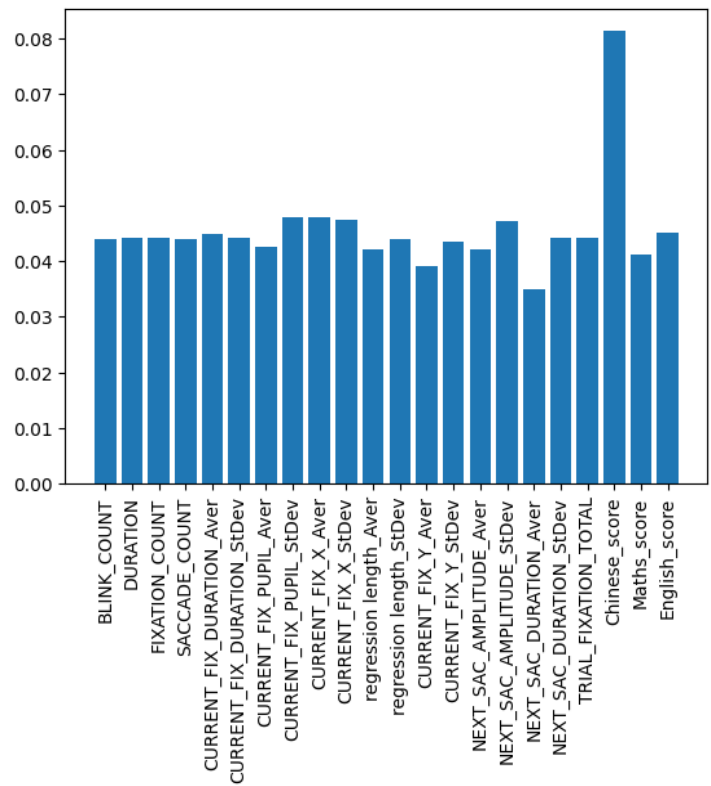

Figure 5 Feature impacts on the accuracy of the proposed method

From Figure 5, it can be found that among the 22 features [18], the feature 'Chinese score' of university entrance examination has the largest influence on the accuracy of the proposed method. And the rest features render similar importance to the accuracy of the proposed method.

Next, we tested the effect of LSTM-based vector embedding as following: the accuracy of the proposed method with different number $k$ of neighboring subjects in LSTM is recorded and illustrated in Figure 6. Note that in case $k=1$, each subject in training data is regarded independently and LSTM is degraded as an ordinary encoder without the memory of 'history'. And in case $k=60$, all subjects in training data are considered jointly for parameter estimation of linear regression.

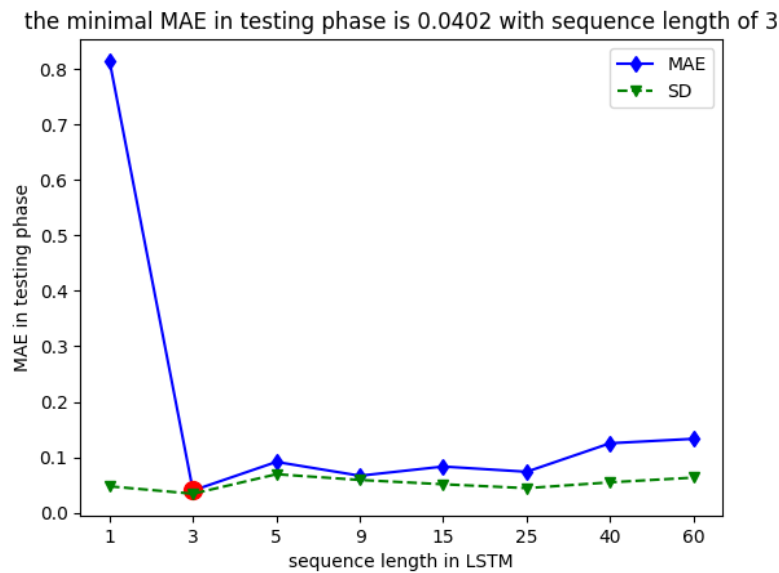

Figure 6 The accuracy of the proposed method with different sequence number $k$ in LSTM

From Figure 6, we can find that the optimal number $k$ of neighboring subjects in LSTM is is neither 1 (each independent subject) nor 60 (all subjects jointly in training). Instead, a relatively small number (e.g., $k=3$) of subjects might be superior to other cases (e.g., $k=1$ or $k=60$). This indicates that the proposed method adopts a proper way of exploiting data dependencies for vector embedding, different from previous methods that either use independent data [15-17] or employs all training data as a whole [18].

Finally, we compared our method with previous methods on score prediction of reading ability, as is shown in Table 3. Although the previous methods and the proposed method all used linear regression for score prediction, the ways of their estimating the parameters for linear regression were different. [16-17] used a classic method of least squares for parameter estimation, [15] employed a Bayesian framework for parameter estimation, and [18] presented a MTJL model for parameter estimation. We adopted two public python classes, sklearn.linear_model.LinearRegression and sklearn.linear_model.BayesianRidge, to implement the method of least squares and Bayesian framework, respectively. All these methods in [15-18] were tested on the same dataset and employed the same 22-dimensional features as those in this paper. Besides, the method in [17] used 4-fold cross-validation at participant level, the methods in [15-16] employed leave-one-out cross-validation, which are listed in Table 3.

Table 3 Comparison among different score prediction methods

| Methods of score prediction | MAE | SD |
|---|---|---|
| Least squares with 4-fold cross-validation [17] | 12.04% | 0.0618 |
| Least squares with leave-one-out cross-validation [16] | 16.90% | 0.0851 |
| Bayesian framework with leave-one-out cross-validation [15] | 15.42% | 0.0759 |
| MTJL Model [18] | 4.91% | 0.9400 |
| **Proposed method** | **4.02%** | 0.0344 |

As can be seen in Table 3, the proposed method achieved best accuracy tested on the same data compared to previous methods, showing its effectiveness for score prediction of reading ability. That is, using episodic learning strategy, the LSTM-based vector embedding combined with light-weighted neural networks exhibited rather good performance on the score prediction of reading ability. Unlike previous methods [15-18] which used manually designed formulas for estimating parameters of linear regression, the proposed methods adopted an 'intelligent' way of the parameter estimation for linear regression by using light-weighted neural networks, indicating the potential application of deep learning methods to the problem of score prediction of reading ability.

## 5. Conclusions

A LSTM-based few-shot learning method for score prediction of reading ability using eye-tracking data was proposed. In the method, LSTM was employed to embed the raw eye-tracking vectors into new vector space, and light-weighted neural networks were utilized to estimate the parameters of linear regression. In addition, few-shot learning strategy was applied to training the proposed regression model on a small-size dataset. The proposed method showed much effectiveness compared to previous methods.

In future works, the proposed method may be improved from the following three aspects: Firstly, the time information of eye-tracking data could be involved with the currently used feature vectors for more precise score prediction. Secondly, generative models for producing synthetic eye-tracking data might be considered to enlarge the present dataset. Thirdly, more complex regression models (e.g., deeper neural networks) would be designed given sufficient training data.

**References:**

[1] Boris Nerušil, Jaroslav Polec, Juraj Škunda, Juraj Kačur. Eye tracking based dyslexia detection using a holistic approach. Sci. Rep., 11(1): 15687. 2021.

[2] Sidney K. D'Mello, Rosy Southwell, Julie Gregg. Machine-learned computational models can enhance the study of text and discourse: A case study using eye tracking to model reading comprehension. Discourse Processes, 57 (5-6): 420-440. 2020.

[3] Patrick Haller, Andreas Säuberli, Sarah E. Kiener, Jinger Pan, Ming Yan, Lena A. Jäger. Eye-tracking based classification of Mandarin Chinese readers with and without dyslexia using neural sequence models. In Proceedings of the Workshop on Text Simplification, Accessibility, and Readability, Association for Computational Linguistics, Abu Dhabi, UAE: 111-118. 2022.

[4] Alae Eddine El Hmimdi, Zoï Kapoula, Vivien Sainte Fare Garnot. Deep learning-based detection of learning disorders on a large scale dataset of eye movement records. BioMedInformatics, 4: 519-541. 2024.

[5] Ivan Vajs, Vanja Kovič, Tamara Papič, Andrej M. Savič, Milica M. Jankovič. Spatiotemporal eye-tracking feature set for improved recognition of dyslexic reading patterns in children. Sensors, 22: 4900. 2022.

[6] Ivan A. VAJS, Goran S. KVAŠČEV, Tamara M. PAPIĆ, Milica M. JANKOVIĆ. Eye-tracking image encoding: Autoencoders for the crossing of language boundaries in developmental dyslexia detection. IEEE Access, 11:3024-3033. 2023.

[7] Stephen Hutt, Aaron Wong, Alexandra Papoutsaki, Ryan S. Baker, Joshua I. Gold, Caitlin Mills. Webcam-based eye tracking to detect mind wandering and comprehension errors. Behavior Research Methods, 56:1-17. 2024.

[8] F. R. Iaconis, M. A. Trujillo Jiménez, G. Gasaneo, O. A. Rosso, C. A. Delrieux. Ordinal pattern transition networks in eye tracking reading signals. Chaos, 33 (5): 053101. 2023.

[9] Harshitha Nagarajan, Vishnu Sai Inakollu, Punitha Vancha, Amudha J. Detection of reading impairment from eye-gaze behaviour using reinforcement learning. Procedia Computer Science, 218: 2734- 2743. 2023.

[10] Yongseok Yoo. Diagnosing Reading Disorders based on Eye Movements during Natural Reading. Journal of Information and Communication convergence engineering, 21 (4): 281- 286. 2023.

[11] Paul Prasse, David R. Reich, Silvia Makowski, Tobias Scheffer, Lena A. Jäger. Improving cognitive-state analysis from eye gaze with synthetic eye-movement data.

Computers and Graphics, 119, 103901. 2024.

[12] Xiangyao Qi, Qi Lu, Wentao Pan, Yingying Zhao, Rui Zhu, Mingzhi Dong, Yuhu Chang, Qin Lv, Robert P. Dick, Fan Yang, Tun Lu, Ning Gu, and Li Shang. CASES: A Cognition-Aware Smart Eyewear System for Understanding How People Read. Proceedings of the ACM on Interactive, Mobile, Wearable and Ubiquitous Technologies, New York, NY, USA, vol.7, no.3. 2023.

[13] Weiqing Shi, Xin Jiang. Predicting Chinese reading proficiency based on eye movement features and machine learning. Reading and Writing. 2024.

[14] Ziqing Xia, Shuhui Lyu, Chun-Hsien Chen, Bufan Liu. An interpretable English reading proficiency detection model in an online learning environment: A study based on eye movement. Learning and Individual Differences, 109: 102407:2024.

[15] Diane C. Mézière, Lili Yu, Erik D. Reichle, Titus von der Malsburg, Genevieve Mcarthur. Using eye-tracking measures to predict reading comprehension. Reading Research Quarterly, 58 (3): 425-449. 2023.

[16] Frans van der Sluis, Egon L. van den Broek. Feedback beyond accuracy: Using eye-tracking to detect comprehensibility and interest during reading. Journal of the Association for Information Science and Technology, 74(1): 3-16. 2023.

[17] Rosy Southwell, Julie Gregg, Robert Bixler, Sidney K. D'Mello. What eye movements reveal about later comprehension of long connected texts. Cognitive Science, 44 (10): 1-24. 2020.

[18] Zehui Zhan, Lei Zhang, Hu Mei, Patrick S. W. Fong. Online learners' reading ability detection based on eye-tracking sensors. Sensors, 16: 1457. 2016.

[19] Sepp Hochreiter; Jürgen Schmidhuber. Long short-term memory. Neural Computation, 9 (8): 1735 – 1780. 1997.

[20] Archit Parnami and Minwoo Lee. Learning from a few examples: A summary of approaches to few-shot learning. arXiv preprint, arXiv:2203.04291. 2022.

[21] Wenqi Ren, Yang Tang, Qiyu Sun, Chaoqiang Zhao, Qing-Long Han. Visual semantic segmentation based on few/zero-shot learning: An overview. IEEE/CAA Journal of Automatica Sinica, 11 (5): 1106-1126. 2024.

[22] Wenbin Li, Ziyi Wang, Xuesong Yang, Chuanqi Dong, Pinzhuo Tian, Tiexin Qin, Jing Huo, Yinghuan Shi, Lei Wang, Yang Gao, Jiebo Luo. LibFewShot: A comprehensive library for few-shot learning. IEEE Transactions on Pattern Analysis and Machine Intelligence, 45: (12): 14938-14955. 2023.

[23] Jiaoyan Chen, Yuxia Geng, Zhuo Chen, Jeff Z. Pan, Yuan He, Wen Zhang, Ian Horrocks, Huajun Chen. Zero-shot and few-shot learning with knowledge graphs: A comprehensive survey. In: Proceedings of the IEEE, 111 (6): 653-685. 2023.

[24] Yisheng Song, Ting Wang, Puyu Cai, Subrota K. Mondal, Jyoti Prakash Sahoo. A

Comprehensive Survey of Few-shot Learning: Evolution, Applications, Challenges, and Opportunities. ACM Computing Surveys, 55 (13s): 1-40. 2023.

[25] Mateusz Ochal, Massimiliano Patacchiola, Jose Vazquez, Amos Storkey, Sen Wang. Few-shot learning with class imbalance. IEEE Transactions on Artificial Intelligence, 4 (5): 1348 – 1358. 2023.

[26] Tomoharu Iwata, Yusuke Tanaka. Few-shot learning for spatial regression. arXiv preprint, arXiv:2010.04360. 2020.

[27] Ning Gao, Hanna Ziesche, Ngo Anh Vien, Michael Volpp. Gerhard Neumann. What Matters for Meta-Learning Vision Regression Tasks? In: Proceedings of the IEEE/CVF Conference on Computer Vision and Pattern Recognition (CVPR): 14776-14786. 2022.

[28] Sungyong Baik, Janghoon Choi, Heewon Kim, Dohee Cho, Jaesik Min, Kyoung Mu Lee. Meta-learning with task-adaptive loss function for few-shot learning. In: Proceedings of the IEEE/CVF International Conference on Computer Vision (ICCV): 9465-9474. 2021.

[29] Kuilin Chen, Chi-Guhn Lee. Incremental few-shot learning via vector quantization in deep embedded space. In: International conference on learning representations (ICLR). 2021.

[30] Pinzhuo Tian, Wenbin Li, Yang Gao. Consistent meta-regularization for better meta-knowledge in few-shot learning. IEEE Transactions on Neural Networks and Learning Systems, 33 (12): 7277-7288. 2022.

[31] Wahyu Fadli Satrya, Ji-Hoon Yun. Combining Model-Agnostic Meta-Learning and Transfer Learning for Regression. Sensors, 23(2):583. 2023.

[32] Sungyong Baik, Myungsub Choi, Janghoon Choi, Heewon Kim, Kyoung Mu Lee. Learning to learn task-adaptive hyperparameters for few-shot learning. IEEE Transactions on Pattern Analysis and Machine Intelligence, 46 (3): 1441 - 1454. 2024.

[33] Minghong Yang, Qinli Yang, Junming Shao, Guoqing Wang, Wei Zhang. A new few-shot learning model for runoff prediction: Demonstration in two data scarce regions. Environmental Modelling & Software, 162: 105659. 2023.

[34] Mike Huisman, Thomas M. Moerland, Aske Plaat, Jan N. van Rijn. Are LSTMs good few-shot learners? Machine Learning, 112 : 4635–4662. 2023.

[35] Michael Phi. Illustrated guide to LSTM's and GRU's: A step by step explanation. Towards Data Science, https://towardsdatascience.com/illustrated-guide-to-lstms-and-gru-s-a-step-by-step-explanation-44e9eb85bf21. 2018.

[36] Da Li, Jianshu Zhang, Yongxin Yang, Cong Liu, Yi-Zhe Song, Timothy M. Hospedales. Episodic training for domain generalization. Proceedings of the IEEE/CVF International Conference on Computer Vision (ICCV): 1446- 1455. 2019.

[37] Chenxin Li, Xin Lin, Yijin Mao, Wei Lin, Qi Qi, Xinghao Ding, Yue Huang, Dong

Liang, Yizhou Yu. Domain generalization on medical imaging classification using episodic training with task augmentation. Computers in Biology and Medicine, 141: 105144. 2022.

[38] Yuqi Cui, Yifan Xu, Dongrui Wu. EEG-based driver drowsiness estimation using feature weighted episodic training. IEEE Transactions on Neural Systems and Rehabilitation Engineering, 27 (11): 2263 - 2273. 2019.

[39] Scott M. Lundberg, Su-In Lee. A Unified Approach to Interpreting Model Predictions. Advances in Neural Information Processing Systems 30 (NIPS). 2017.

[40] Simon Meyer Lauritsen, Mads Kristensen, Mathias Vassard Olsen, Morten Skaarup Larsen, Katrine Meyer Lauritsen, Marianne Johansson Jørgensen, Jeppe Lange, Bo Thiesson. Explainable artificial intelligence model to predict acute critical illness from electronic health records. Nature Communications, 11: 3852. 2020.